\documentclass{elsart}

\usepackage{graphicx}
\usepackage{amssymb}

\begin{document}

\begin{frontmatter}



\title{The hexatic phase \\of the two-dimensional hard disk system}


\author{A.\ Jaster}

\address{Innovista, Physics Department, Marienstra\ss e 6, 30171 Hannover, Germany}

\begin{abstract}
We report Monte Carlo results for the two-dimensional hard disk
system in the transition region. Simulations were performed in the
$NVT$ ensemble with up to $1024^2$ disks. The scaling behaviour of
the positional and bond-orientational order parameter as well as
the positional correlation length prove  the existence of a
hexatic phase as predicted by the
Kosterlitz-Thouless-Halperin-Nelson-Young theory. The analysis of
the pressure shows that this phase is outside a possible
first-order transition.
\end{abstract}

\end{frontmatter}

The nature of the two-dimensional melting transition has been an
unsolved problem for many years~\cite{STRAND,GLACLA}. The
Kosterlitz-Thouless-Halperin-Nelson-Young (KTHNY) 
theory~\cite{KTHNY1,KTHNY2,KTHNY3,KTHNY4} predicts two continuous
transitions. The first transition occurs when the solid
(quasi-long-range positional order, long-range orientational
order) undergoes a dislocation unbinding transition to the hexatic
phase (short-range positional order, quasi-long-range
orientational order). The second transition is the disclination
unbinding transition which transforms this hexatic phase into an
isotropic phase (short-range positional and orientational order).
An alternative scenario has been proposed by 
Chui~\cite{CHUI1,CHUI2}. He presented a theory via spontaneous
generation of grain boundaries, i.e.\ collective excitations of
dislocations. He found that grain boundaries may be generated
before the dislocations unbind if the core energy of dislocations
is sufficiently small, and predicted a first-order transition.
This is characterized by a coexistence region of the solid and
isotropic phase, while no hexatic phase exists. Another proposal
was given by Glaser and Clark~\cite{GLACLA}. They considered a
detailed theory where the transition is handled as a condensation
of localized, thermally generated geometrical defects and found
also a first-order transition. Calculations based on the
density-functional approach were done by Ryzhov and 
Tareyeva~\cite{RYZTAR}. They derived that the hexatic phase cannot exist in
the hard disk system.

Even for the simple hard disk system no consensus about the
existence of a hexatic phase has been established thus far. The
melting transition of the hard disk system was first seen in a
computer simulation by Alder and Wainwright~\cite{ALDWAI}. They
used a system of 870 disks and molecular dynamics methods ($NVE$
ensemble) and found that this system undergoes a first-order phase
transition. But the results of such small systems are affected by
large finite-size effects.  Simulations performed in the last
years used Monte Carlo (MC) techniques either with constant volume
($NVT$ ensemble)~\cite{ZOCHLE,ZOLCHE,WEMABI1,WEMABI2,MIWEMA,SENIBI,BISENI,JASTER21,JASTER22,JASTER41,JASTER42}
or constant pressure ($NpT$ ensemble)~\cite{LEESTR,FEALST1,FEALST2}. 
Zollweg, Chester and Leung~\cite{ZOCHLE} made detailed investigations of large systems up to
$16384$ particles, but draw no conclusives about the order of the
phase transition. The analysis of Zollweg and Chester~\cite{ZOLCHE} 
for the pressure gave an upper limit for a
first-order phase transition, but is compatible with all other
scenarios. Lee and Strandburg~\cite{LEESTR} used isobaric MC
simulations and found a double-peaked structure in the volume
distribution. Lee-Kosterlitz scaling led them to conclude that the
phase transition is of first-order. However, the data are not in
the scaling region, since their largest system contained only 400
particles. MC investigations of the bond-orientational order
parameter via finite-size scaling (FSS) with the block analysis
technique of 16384 particle systems were done by Weber, Marx and
Binder~\cite{WEMABI1,WEMABI2}. They also favoured a first-order
phase transition. In contrast to this, Fern\'{a}ndez, Alonso and
Stankiewicz~\cite{FEALST1,FEALST2}\footnote{For a critical
discussion of this work see Ref.~\cite{WMFAS1,WMFAS2}.} predicted
a one-stage continuous melting transition, i.e.\ a scenario with a
single continuous transition and consequently without a hexatic
phase. Their conclusions were based on the examination of the
bond-orientational order parameter of different systems up to
15876 particles and hard-crystalline wall boundary conditions.
Mitus, Weber and Marx~\cite{MIWEMA} studied the local structure of
a system with $4096$ hard disks. From the linear behaviour of a
local order parameter they derived bounds for a possible
coexistence region. Sengupta, Nielaba and Binder~\cite{SENIBI}
simulated a dislocation free triangular solid of hard disks using
a constrained Monte Carlo algorithm and showed that a KTHNY
transition preempts a first-order transition. Combining
renormalisation groups ideas with MC input they derived also an
estimate of $\rho_{\rm m}=0.914(2)$ for a possible
hexatic-to-crystal transition~\cite{BISENI}. Finally, in~\cite{JASTER21,JASTER22}
the present author determined the
disclination binding transition density $\rho_{\rm i}=0.899(1)$
from measurements of the bond-orientational correlation length and
susceptibility in the isotropic phase as well as the critical
exponent $\eta_6=0.25(4)$ from finite-size scaling of systems with
up to 16384 particles. The results are in agreement with the KTHNY
theory, while a first-order phase transition with small
correlation length and a  one-stage continuous transition can be
ruled out. By studying short-time behaviour and FSS of the
positional order~\cite{JASTER41,JASTER42} we also estimated the
dislocation binding density with $\rho_{\rm m} \approx 0.933$ as
well as the critical exponent $\eta \approx 0.200$.

In this letter, we present results obtained through MC simulations
in the $NVT$ ensemble to answer the question of the existence of a
hexatic phase and therefore the kind of the phase transition.
Although we have already shown that the disclination binding
density $\rho_{\rm i}=0.899(1)$ is lower than the dislocation
binding density $\rho_{\rm m} \approx 0.933$, no direct
observation of the hexatic phase so far exists. Therefore, we
examine the positional order as well as the orientational order
within this region at $\rho = 0.914$ and $\rho = 0.918$.
Simulations in the $NVT$ ensemble are computationally less expensive 
than in the $NpT$ ensemble. The density region is outside a 
possible first order transition thus there is no necessity 
to perform constant pressure simulations.

We consider systems of $N=32^2$ up to  $1024^2$ hard disks in a
two-dimensional rectangular box with ratio $\sqrt{3}:2$. 
Additionally, we studied three systems in a square box to determine
the influence of boundary conditions. The disk
diameter is set equal to one.  For the simulations we used an
improved updating scheme~\cite{JASTER1}, in which the conventional
Metropolis step of a single particle is replaced by a collective
(non-local) step of a chain of particles. 

Due to the improved updating
scheme, which reduces autocorrelation times, we were able to
perform simulations of large systems within the transition region.
Although, the scaling behaviour is probably not changed compared
to the Metropolis algorithm, the chain algorithm significantly speeds up 
the simulations. Especially in the hexatic phase with 
large positional order length, this algorithm is
advantageous  \footnote{The chains or particles moved in a single
step are much longer than those observed for 
$\rho=0.898$ \cite{JASTER1}.}.

We measured the $k$th moment of the global bond-orientational
order parameter
\begin{equation}
{\psi_6}^k= \left | \frac{1}{N} \sum_{i=1}^N \frac{1}{N_i} \sum_j
\exp \left (6\, {\rm i} \, \theta_{ij} \right ) \right |^k \ ,
\end{equation}
where the sum on $j$ is over the $N_i$ neighbours of this particle
and $\theta_{ij}$ is the angle  between the particles $i$ and $j$
and an arbitrary but fixed reference axis. Two particles are
defined as neighbours if the distance is less than $1.4 a$, where
$a=\sqrt{2/\sqrt{3}\rho}$ is the average lattice spacing of a
perfect crystal. This definition is computationally less expensive
than precise determination by the Voronoi construction. While the
values of $\psi_6$ shown in table~\ref{table1} depend on the
definition of neighbours, the conclusions are independent of the
method used.

The $k$th moment of the positional order parameter is defined as
\begin{equation}
{\psi_{\rm pos}}^k= \left | \frac{1}{N} \sum_{i=1}^N  \exp ( {\rm
i} \,{\rm \bf G} \cdot {\rm \bf r}_i )  \right |^k \ ,
\end{equation}
where ${\rm \bf G}$ is a reciprocal lattice vector and ${\rm \bf
r}_i$ denotes the position of particle $i$. The magnitude of ${\rm
\bf G}$ is given by $2\pi/a$, while the orientation was determined
from the global bond-orientation. With $\psi_6 \sim \exp(6\,{\rm
i}\,\theta)$ we defined the angle $\theta$ ($-\pi/6\le\theta<\pi/6$)
and therefore the orientation of the crystal and ${\rm \bf G}$.
The fourth-order cumulant for the positional order is given by
\begin{equation}
\label{Upos} U_{\rm pos}=1-\frac{ \left \langle {\psi_{\rm pos}}^4
\right \rangle}{3 \left \langle {\psi_{\rm pos}}^2  \right \rangle
^2} \ .
\end{equation}
The bond-orientational correlation length $\xi_{\rm pos}$ was
extracted from the `zero-momentum' correlation function of
$\psi_{\rm pos}({\rm \bf x})$
\begin{equation}
g_{\rm pos}(x) =  \left ( \frac{1}{N_k} \sum_y \psi_{\rm pos}(x,y)
\right ) ^* \, \left ( \frac{1}{N_k'} \sum_{y'} \psi_{\rm
pos}(0,y') \right )  \ ,
\end{equation}
by fitting the data with a single $\cosh$, where $N_k$ denotes the
number of particles in a stripe between $x+\Delta x/2$ and
$x-\Delta x/2$~\cite{JASTER22}. The pressure is calculated from
the pair correlation function $g(r)$
\begin{equation}
\label{pressure} \frac{p A_0}{NkT} = \frac{\sqrt{3}}{2} \rho \left
( 1+ \frac{\pi}{2}\rho \, g(1) \right )\ ,
\end{equation}
where $A_0$ is the close-packed area of the system, i.e.\ $A_0=N \sqrt{3}/2$.
%%%%%%%%%%%%%%%%%%%%%%%%%%%%%%%%%%%%%%%%%%%%%%%%%%%%%%%%%%%%%%%%%%%%%%%%
\begin{table*}
\begin{center}
\caption{ \label{table1} The density of three and eight
coordinated particles, the pressure, the second moment of the
global bond-orientational and positional order parameter, the
positional fourth-order cumulant and the positional correlation
length for different densities,  boundary conditions and system sizes. }
%\begin{ruledtabular}
%\begin{tabular}{ccrddD{.}{.}{1.7}D{.}{.}{1.9}D{.}{.}{1.7}D{.}{.}{1.7}D{(}{(}{4}}
\begin{tabular}{ccrllllllr}
\hline
\hline
$\rho$ & box & $N$ & 
\multicolumn{1}{c}{$n_3 \cdot 10^{6}$} & 
\multicolumn{1}{c}{$n_8 \cdot 10^{5}$} &
\multicolumn{1}{c}{$\frac{pA_0}{NKT}$}  &
\multicolumn{1}{c}{${\psi_6}^2$} &
\multicolumn{1}{c}{${\psi_{\rm pos}}^2$} &
\multicolumn{1}{c}{$U_{\rm pos}$} &
\multicolumn{1}{c}{$\xi_{\rm pos}$}  \\
\hline
$0.914$   & rect.  &  $128^2$ &  3.76(15)  &  4.33(15)   &  7.921(6)     &   0.5314(20)   &   0.15(3)    &  0.60(3)   & 47(6)     \\
$0.914$   & rect.  & $256^2$  &  4.29(4)   &  5.11(4)    &  7.968(3)     &   0.5181(10)   &   0.017(3)   &  0.55(2) &  21(3)    \\
$0.914$   & rect.  & $512^2$ &  4.34(4)    &  5.21(4)    &  7.975(3)     &   0.5126(30)   &  0.0026(8)   &  0.53(2)  &  15(3)    \\
$0.918$   & rect.  & $32^2$   &  0.7(4)    &  1.3(4)     &  7.947(12)    &   0.5711(30)   &   0.41(4)    &  0.66(1)   &  17(10)   \\
$0.918$   & rect.  & $64^2$   &  3.0(2)    &  3.0(1)     &  8.022(5)     &   0.5616(10)   &   0.33(4)    &  0.65(1)   &  37(4)     \\
$0.918$   & rect.  & $128^2$  &  3.4(2)    &  3.3(1)     &  8.050(4)     &   0.5558(12)   &   0.147(12)  &  0.64(1)   &  49(4)     \\
$0.918$   & sq.    &  $128^2$  &  2.8(1)    &  2.8(1)    &  8.019(5)     &    0.5612(9)   &   0.183(8)   &  0.65(1)   &  52(10)     \\
$0.918$   & rect.  & $256^2$  &  2.7(3)    &  2.9(2)     &  8.021(10)    &   0.5608(16)   &   0.117(25)  &  0.64(1)   & 110(35)     \\
$0.918$   & sq.    & $256^2$  &  3.0(1)    &  3.0(1)     &  8.029(5)    &   0.5592(20)   &   0.087(20)  &  0.65(1)   & 88(30)     \\
$0.918$   & rect.  & $512^2$  &  3.1(1)    &  3.1(1)     &  8.034(3)     &   0.5579(10)   &   0.0095(20) &  0.55(3)   &  32(8)    \\
$0.918$   & sq.    & $512^2$  &  3.0(1)    &  3.1(1)     &  8.038(5)    &   0.5563(14)   &   0.0061(17) &  0.55(4)  & 32(6)     \\
$0.918$   & rect.  & $1024^2$ &  3.1(1)    &  3.2(1)     &  8.041(7)     &   0.5561(15)   &   0.0020(3)  &  0.54(3)  &  37(4)    \\
\hline
\hline
\end{tabular}
%\end{ruledtabular}
\end{center}
\end{table*}
%%%%%%%%%%%%%%%%%%%%%%%%%%%%%%%%%%%%%%%%%%%%%%%%%%%%%%%%%%%%%%%%%%%%%%%%

%%%%%%%%%%%%%%%%%%%%%%%%%%%%%%%%%%%%%%%%%%%%%%%%%%%%%%%%%%%%%%%%%%%%%%%%
\begin{table}
\begin{center}
\caption{ \label{table2} Estimated integrated autocorrelation time 
measured in units of thousand sweeps of the chain Metropolis algorithm.
We used different (optimized) step sizes for each density.}
%\begin{ruledtabular}
\begin{tabular}{cccccccc}
\hline
\hline
$\rho$ & box & $N$ &
\multicolumn{1}{c}{$n_3$} & 
\multicolumn{1}{c}{$n_8$} &
\multicolumn{1}{c}{$\frac{pA_0}{NKT}$}  &
\multicolumn{1}{c}{${\psi_6}^2$} &
\multicolumn{1}{c}{${\psi_{\rm pos}}^2$} \\
\hline
$0.914$  & rect. & $256^2$   &  $<$1    &  $<$1    &   $<$1   &   32   &   54       \\
$0.918$  & rect. & $512^2$   &  $<$1    &  $<$1    &   $<$1   &   59   &   270        \\
\hline
\hline
\end{tabular}
%\end{ruledtabular}
\end{center}
\end{table}
%%%%%%%%%%%%%%%%%%%%%%%%%%%%%%%%%%%%%%%%%%%%%%%%%%%%%%%%%%%%%%%%%%%%%%%%
All simulations of systems in the rectangular box started from 
the ordered state while we used a closed-packed state as
initial condition in case of the square box \footnote{Disordered initial states 
for the hard disk model at high densities  are
more difficult to simulate and equilibration is  more time consuming.}.
Careful attention has been paid to the equilibration of the
systems.
For that purpose we estimated autocorrelation
times from binning \footnote{We
built blocks of subsequent configurations, called bins, and averaged the 
quantities first in the bin. The obtained bin averages themselves 
can be considered as the results of single measurements. If the bins are 
large enough, then the average values in different bins are practically
uncorrelated  and the obtained statistical errors are correct. 
All given statistical errors are obtained in this way, i.e.\ taking 
correlations into account.
This is also a way to estimate the autocorrelation time.}. Additionally,  
two systems were studied in detail,
i.e.\ we monitored all quantities and measured the autocorrelation times. 
We used the defect density as an estimate for the equilibration  of 
dislocations. The results are given
in table~\ref{table2}.  The autocorrelation times are much smaller than 
the number of warm-up sweeps, which were at least one million.
The number of measurement sweeps ranged from 2 to 20 millions.

For a hexatic phase --- as predicted by the KTHNY theory --- FSS
implies ${\psi_6}^2 \sim L^{-\eta_6}$ for the orientational order
parameter, while the positional order parameter should scale with
${\psi_{\rm pos}}^2 \sim L^{-2}$ for large enough system sizes ($L
\gg \xi_{\rm pos}$). The cumulant $U_{\rm pos}$ should decrease
with increasing system size $L$ for $\rho < \rho_{\rm m}$.

We measure ${\psi_6}^2$, ${\psi_{\rm pos}}^2$, $U_{\rm pos}$,
$\xi_{\rm pos}$ and $pA_0/NKT$ as well as the density of three
($n_3$) and eight coordinated particles ($n_8$) \footnote{These
values are obtained from the definition for neighbours given
above.}. We use systems of $N=32^2, 64^2, 128^2, 256^2, 512^2$ and
$1024^2$ particles. All results are given in table~\ref{table1}.
Statistical errors have been calculated by binning. We also added 
systematic errors coming from the interpolation to $g(1)$ for the
pressure as well as errors coming from the interval used for fitting in
case of the correlation length. Simulations of the hard disk system are
not only affected by usual finite-size effects, but also by
systematic errors coming from the boundary conditions. Even for a
rectangular box of ratio $\sqrt{3}:2$ and (quasi-) long-range
order crystal tilting occurs. This leads to large autocorrelation
times as well as additional, complicated finite-size effects
compared to simple lattice models.

\begin{figure}
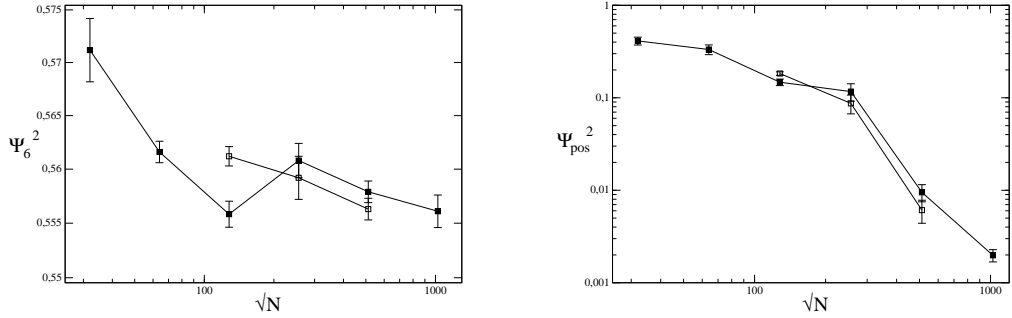

\begin{center}
\includegraphics[width=6.0cm]{fig1a}
\hspace{1cm}
\includegraphics[width=6.0cm]{fig1b}
\caption{\label{fig_psi} Double logarithmic plots of the second
moment of the bond-orientational (left figure) and the positional
order (right figure) parameter at $\rho=0.918$ as a function of $\sqrt{N}$, which
is proportional to the system size $L$. 
Full symbols are for the rectangular box while open
symbols denote square boxes.
The lines are guides for the eye.}
\end{center}
\end{figure}

The pressure and positional correlation length at $\rho=0.918$ are nearly
independent for system sizes with $N \ge 64$. The results of
$n_3$ and $n_8$ for $N=32^2$ show the suppression of defects for
systems that are too small. Leaving out the data for $N=32^2$ 
the results  at $\rho=0.918$ seem to be mostly consistent. 
The results for ${\psi_6}^2$ and
${\psi_{\rm pos}}^2$ are shown in  fig.~\ref{fig_psi}. Obviously,
the orientational order is (quasi-) long-range, while the
positional order is short-ranged \footnote{We carefully 
tried to equilibrate the systems. The measurements of the correlation time
shows that the only critical point is $N=1024^2$. However, 
even a too short equilibration
time for the largest system wouldn't affect our conclusions.
The reason is that due to the initial ordered state for
the rectangular boxes, a non-equilibrated system would 
lead to too high values for the order 
parameter, i.e.\ the decrease of ${\psi_{\rm pos}}^2$ for increasing system sizes
would be even larger.} 
which is confirmed by the
decrease of $U_{\rm pos}$. Deviations are
caused by statistical errors coming from large
autocorrelation times and systematic errors due to boundary
conditions. For example, the increase of ${\psi_6}^2$ 
for lower system sizes ($N=32^2$) and rectangular
boundary conditions is caused by stabilization effects.
The (quasi-) long-range of the orientational order 
is consistent with previous measurements~\cite{JASTER22} where 
$\rho_{\rm i}=0.899(1)$ was determined.
The measurements at $\rho=0.914$  confirm the short-range 
positional order in this range.

Although, the results should be independent from the initial 
conditions, our data can differ due to the different boundary
conditions used.
However, the data should converge for increasing system sizes.
Also, for larger systems the correlation length, which is 
direction dependent for smaller
systems, should be isotropic.

%%%%%%%%%%%%%%%%%%%%%%%%%%%%%%%%%%%%%%%%%%%%%%%%%%%%%%%%%%%%%%%%%%%%%%%%
\begin{figure}
\begin{center}
\includegraphics[width=6.0cm,height=5.2cm]{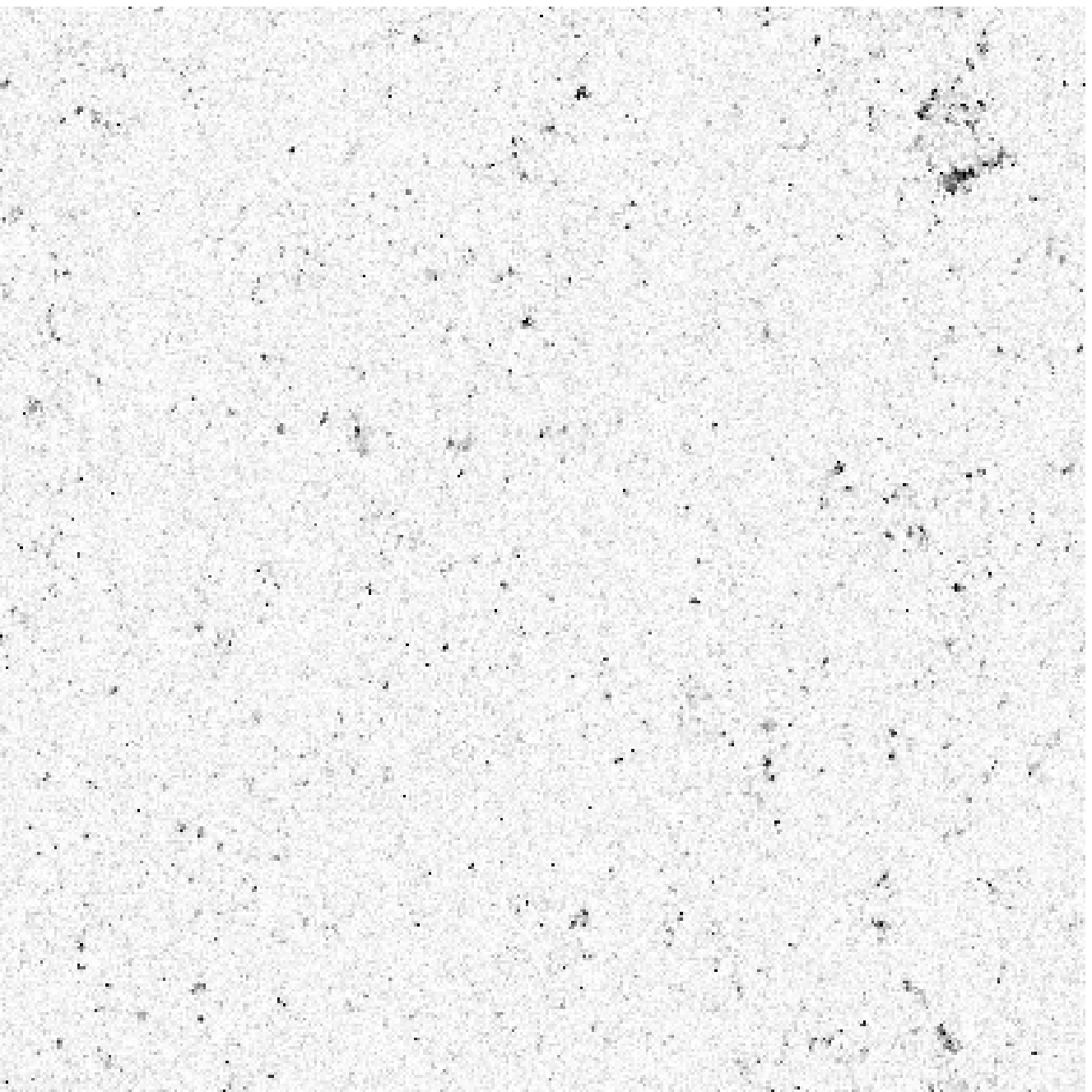}
\hspace{1cm}
\includegraphics[width=6.0cm,height=5.2cm]{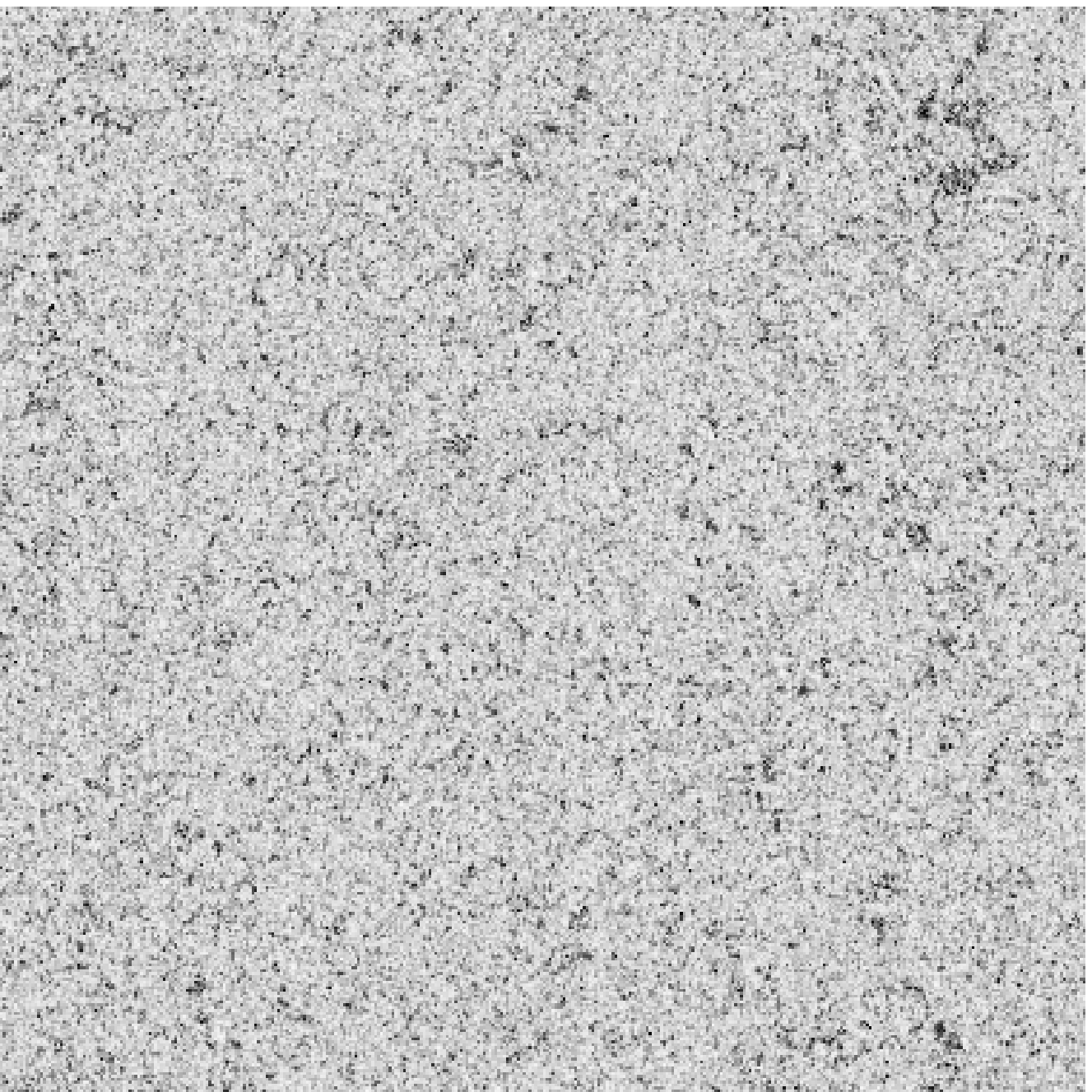}
\caption{\label{fig_order} Visualisations of the
local bond-orientational order parameter ${\psi_{\rm 6}}$ of a typical configuration
with $1024^2$ particles in a rectangular box. In the left picture the amount of black represents
the local orientation. White areas correspond to the inital orientation
of the perfect ordered system ($\theta=0$). The larger the absolute value of $\theta$,
the darker the areas. The right picture shows the degree of order, i.e.\
the local value of $0 \le {\psi_{\rm 6}}^2 \le 1$ is visualised. White areas 
correspond to perfect ordered areas. }
\end{center}
\end{figure}

%%%%%%%%%%%%%%%%%%%%%%%%%%%%%%%%%%%%%%%%%%%%%%%%%%%%%%%%%%%%%%%%%%%%%%%%
A visualisation of the orientational order for a
typical configuration at $\rho=0.918$ with $N=1024^2$ particles
in the rectangular box is shown in
fig.~\ref{fig_order}a and b. 
We divided the system into $380^2$
subsystems with approximately 7  particles per subsystem and
calculated ${\psi_6}^k$ averaged over the particles inside.
The left picture visualises the local orientation. 
The initial state of a perfect ordered crystal
($\theta=0$) corresponds to white areas. Areas with $|\theta|>0$
are grey, where the amount of black is proportional
 $|\theta|$. The case $|\theta|= \pi/6$ corresponds to black regions.
The right picture shows the local order, i.e.\ 
the amount of black is chosen proportional to $1-{\psi_6}^2$.
Thus, the initial state of a perfect ordered crystal
corresponds to a white area. The two pictures show that areas
with orientations significant different from $\theta=0$ 
are normally small, i.e.\  have less than 100 particles. 
%%%%%%%%%%%%%%%%%%%%%%%%%%%%%%%%%%%%%%%%%%%%%%%%%%%%%%%%%%%%%%%%%%%%%%%%
%%%%%%%%%%%%%%%%%%%%%%%%%%%%%%%%%%%%%%%%%%%%%%%%%%%%%%%%%%%%%%%%%%%%%%%%

Our simulations of the hard disk model in the $NVT$ ensemble at
$\rho=0.914$ and $\rho=0.918$ prove  the short-range of the
positional order  and 
therefore the existence of a hexatic phase as predicted
by the KTHNY theory. The positional 
correlation length is about $15$ at $\rho=0.914$ and 
$40$ at $\rho=0.918$, respectively. The orientational
order is quasi-long-range. The scaling of ${\psi_6}^2$
yields $\eta_6 \approx 0.015$ at $\rho=0.914$ and 
$\eta_6=0.005(3)$ at $\rho=0.918$, respectively. However, we
cannot rule out long-range orientational order. 
The observed phase is not within a
possible first-order phase transition since the pressure $8.04(1)$ 
at $\rho=0.918$
is higher than that of such a transition $7.95(1)$~\cite{JASTER22}. 
Therefore, a one-stage continuous transition~\cite{FEALST1,FEALST2}
as well as a first-order phase transition
from the isotropic to the solid phase can be ruled out. Taking the
results of previous measurements~\cite{JASTER21,JASTER22} a
KTHNY-like phase transition is most likely. However, a first-order
phase transition from the isotropic to the hexatic phase with very
large orientational order correlation lengths cannot be ruled out.
Detailed investigations of the pressure around $\rho_{\rm i}$
would be necessary to make a decision. Also, the exact value of
$\rho_{\rm m}$ and therefore of $\eta$ as well as the behaviour of
$\eta_6$ --- which should approach zero according to the KTHNY
theory --- have to be examined in order to reach a final
conclusion as to the question of the kind of transition.

\section*{Acknowledgments}
We thank Kurt Binder and Lothar Sch\"{u}lke for helpful comments
on our draft and Innovista in  Hannover for providing computer time. 
We also wish to thank Claudia Rivi\`{e}re for correcting grammatical errors and typos. 
Especially we benefited from discussions with Conny Tollp.\ Deiters.

\bibliography{paper-PLA}
\bibliographystyle{elsart-num}

\end{document}